\documentclass[10pt,aps,prb,twocolumn,showpacs,amssymb,floatfix]{revtex4-2}
\usepackage{amsmath,graphics,epsfig,mathrsfs}
\usepackage{epstopdf}
\usepackage{bm}
\usepackage{color}
\usepackage{soul}
\usepackage{subfigure}
\usepackage{hyperref}
\usepackage{physics}
\usepackage{xcolor}
\usepackage{tikz}
\usetikzlibrary{tikzmark,calc}
\usepackage{floatrow}
\usepackage[compact]{titlesec}
\titlespacing{\section}{0pt}{*3}{*3}
\titlespacing{\subsection}{0pt}{*3}{*3}
\newfloatcommand{capbtabbox}{table}[][\FBwidth]
\hypersetup{backref=true,pagebackref=true,hyperindex=true,colorlinks=true,breaklinks=true,urlcolor=
red,linkcolor=blue,bookmarks=true,bookmarksopen=true,citecolor=red}

\begin{document} 
\title{Direct Scheme Calculation of the Kinetic Energy Functional Derivative Using Machine Learning}
\author{H. Saidaoui$^1$}\email{hamed.saidaoui@kaust.edu.sa}
\author{S. Kais$^2$}
\author{S. Rashkeev$^1$}
\author{FH. Alharbi$^{3,4}$}
\affiliation{$^1$Qatar Environment and Energy Research Institute (QEERI), Hamad Bin Khalifa University, Doha, Qatar \\
$^2$Department of Chemistry, Physics, and Birck Nanotechnology Center, Purdue University \\
$^3$Electrical Engineering Department, KFUPM, Dhahran, Saudi Arabia \\
$^4$K.A.CARE Energy Research \& Innovation Center, Dhahran, Saudi Arabia}

\begin{abstract}	
we report a direct scheme calculation of the kinetic energy functional derivative using Machine Learning. Support Vector Regression and Kernel Ridge Regression techniques were independently employed to estimate the kinetic energy functional and its derivative. Even though the accuracy should have been a decisive factor in modeling a realistic functional, we show that at a certain level it affects the generalizability of the model. By choosing the right regularization term and by considering a reasonable interplay between it and the accuracy, we were able to deduce the functional derivative from a model that was trained to estimate the kinetic energy. Although the derivative calculations demand very high accuracy to account for small variations of the electron density, the developed estimator was capable of capturing these extremely small changes of the electron density. This work pours into highly effective implementation of the orbital-free density functional theory as it employs only direct calculation schemes.
\end{abstract}
\maketitle

\section{Introduction}

Atomic scale calculation has become one of the essential methodologies in nowadays scientific activities. Amongst a large number of its methods and techniques, density functional theory (DFT) appears to be the most popular and versatile quantum mechanical theory in investigating the electronic structure. It was introduced in the seminal theory of Thomas and Fermi \cite{Thomas, Fermi}, had its theoretical foundations been laid by Hohenberg and Kohn \cite{Hohenberg-kohn} and had been made practical by adopting the Kohn-Sham fictitious system scheme \cite{Kohn-sham}.
The Kohn-Sham DFT has witnessed an unprecedented success story in showcasing both the accuracy and the low computational cost. It was successful in describing many material properties and - to a good extent - was an efficient tool in investigating chemical systems. However, DFT suffers when it comes to modeling bulk solids band gaps, Van der Waals interactions and strongly correlated systems \cite{Klimes, Burke_perspectives, Becke_review,Jones_RMP}, in addition to the high computational cost associated with nowadays demanding scientific problems.
In spite of being a direct manifestation of the density-based Hohenberg-Kohn theorems, DFT scheme uses Kohn Sham orbitals in order to calculate densities, rendering the computation prohibitively expensive when investigating systems with large number of electrons N. Meanwhile many attempts have been made in order to tackle these flaws, those trials resulted in ameliorating the overall accuracy by improving the pseudo-potentials and the exchange-correlation (XC) functionals. However much less improvement has been achieved in lowering the computational cost\cite{Burke_dftholes}.\\
\indent To this end, orbital-free density functional theory \cite{Ligneres} (OFDFT) seemed to be a good alternative to the current stagnated orbital-based density functional theory from calculation cost point of view. Nevertheless, its practicality was every time faced by the long-standing issue of its functionals accuracy.
Kinetic energy (KE) is the leading term of an electronic system total energy, therefore, errors made in approximating it have a dramatic impact on the total energy accuracy. Even though considerable improvement has been made in calculating and approximating exchange-correlation functionals \cite{Kohn-sham, Perdew_GGA, Becke, Lee, Becke_1993, Perdew_PBE}, the Kinetic  (Kohn Sahm) energy remains a bottleneck to tackle for a full DFT-orbital free implementation.
The first analytical expressions of the kinetic energy functionals were given by the Thomas Fermi functional 
\begin{equation}\label{eq:Thomas}
T_{TF}[n(\textbf{r})] = c_D \int{n(\textbf{r})^{\frac{D+2}{D}} \, d^Dr}
\end{equation}
 for uniform densities and by the Von Weizsacker functional \cite{Weizsacker}
 \begin{equation}\label{eq:Weizsacker}
T_{vW}[n(\textbf{r})] = \frac{1}{8}\int{\frac{|\nabla(n(\textbf{r}))|^2}{n(\textbf{r})} \, d^Dr}
\end{equation}
\noindent for single orbital systems, respectively. Both functionals are given in a $D$ dimensional space where $c_D$ is a $D$-dependent constant.
Following that, many attempts have been devoted to get an accurate KE functional, ranging from proposing linear combination of the aforementioned functionals to employing conventional gradient expansion and enforcing linear response behavior, these attempts were shiny for some systems but were overall non-transferable. \cite{Acharya, Garcia, Kirzhnits,Hodges, Wang,Smargiassi,Foley,Gal,Sim}\\
\indent Recently, new methods which are based on learning from data, have been proposed to approach the OFDFT \cite{SnyderPRL, SnyderJCP, BypassingDerv, Fahhad2017} from a different perspective. Although an overall good accuracy has been achieved throughout these attempts, the applicability of some of these methods are limited when coming to the functional derivative calculations.\\

\begin{figure}[h!]
  \centering
  \includegraphics[width=9.2cm]{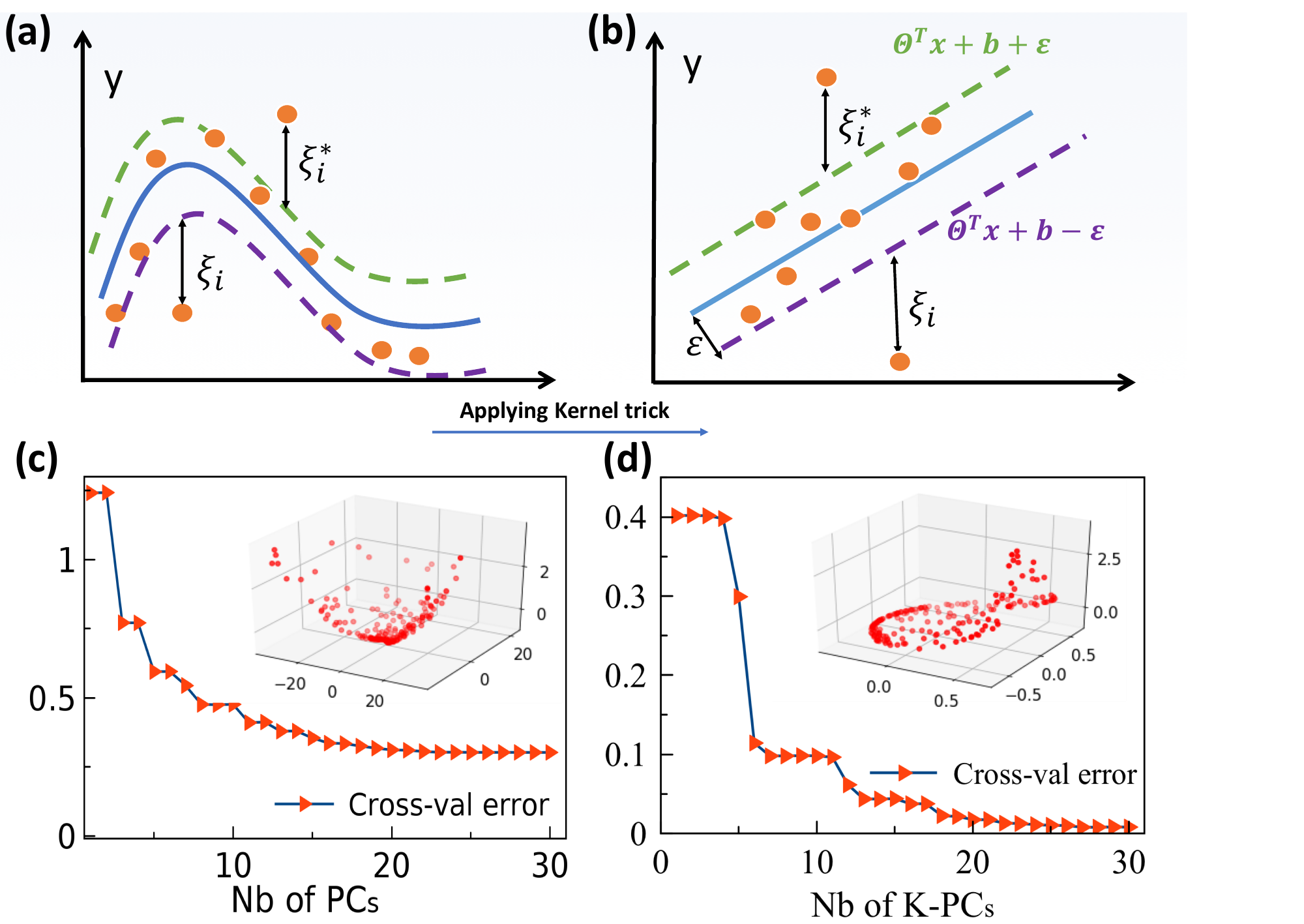}
  \caption{(Color online) Impact of the kernel mapping on the accuracy and speed of the prediction. (a) refers to a fictitious 1-d illustration of a nonlinear function which is transformed to a linear function (b) after the application of the kernel trick. The main plots of (c) and (d) represent the 	cross validation errors as function of the principal components and the kernel principal components, respectively. Insets of (c) and (d) are 3-D plotting of the normalized kinetic energy function of the two first principal components  before and after kernelization, respectively.} \label{fig:fig1}
\end{figure}

\indent In the current work, we develop a method based on Machine Learning (ML) that outputs accurate KE functionals, we then calculate the functional derivative, in a direct manner, without any further training. We use support vector regression (SVR) \cite{VapnikSVM} and kernel ridge regression (KRR) tools throughout this work. This article is organized as follows. We detail the usage of the SVR tool in Sec. \ref{sec:sec2} (we refer the reader to Ref. \cite{SnyderPRL} for a detailed implementation of KRR in predicting KE functionals). In Sec. \ref{sec:sec3}, we discuss the accuracy of the built estimators in predicting KE functionals. Section. \ref{sec:sec4} is devoted to showcase the main finding of this article, namely the direct scheme calculation of the KE functional derivative. Finally, Sec. \ref{sec:sec5} concludes our study. 
 
 \section{Theoretical background}\label{sec:sec2}
 
 As any other ML tool, SVR learns the association between the input vector $\rho$ (referring to the electron density) and the output $y$, i.e, the associated kinetic energy. In order to obtain the training set of densities, we consider the simple system of free particles in a box with a varying potential consisting of a sum of dips randomly centered between two infinite walls:
\begin{equation}\label{eq:potential}
v(x)= \bigg\{ 
\begin{array}{cc}
-\sum^{n_d}_{i=1}a_i \exp(\frac{-(x-b_i)^2}{2c^2_i}) & 0<x<L \\
\infty & \text{otherwise}, \\
\end{array}
\end{equation}
where $a_i$,  $b_i$ and $c_i$ have random values ranging from $\{1-10\}$, $\{0.2-0.8\}$ and $\{0.03-0.1\}$, respectively. $L$ is the well width and $n_d$ is the number of dips. A similar (but much restricted) three dips potential has been used for similar machine learning calculations\cite{SnyderPRL,Fahhad2017}. We use the Galerkin method \cite{Galerkin} to solve the Schrodinger equation associated with each potential. Generated densities are then calculated and sorted in the matrix $\hat{\bm \rho}$. SVR is mainly based on the kernel mapping method which transforms a non-linear problem to a linear one in a higher dimensional space. Here the aim is to find a function that learns from the training dataset and generalizes to new unseen data. The continuous function being approximated can be written as: 
\begin{equation}
 f[\hat{\bm\rho},\hat{\Theta}] = \langle \hat{\Theta}, \hat{\bm\rho} \rangle + b
\end{equation}
where $\langle \hat{\Theta}, \hat{\bm\rho} \rangle$ stands for the inner product between the weight vector $\hat{\Theta}$ and the input vector $\hat{\bm\rho}$, $b$ is the bias term.
$\hat{\bm\rho}$ is an $(l,n_f)$ matrix that contains all the training set densities, $\{\rho_i\}$. $l$ is the dimension of the training set while $n_f$, the number of features, is equal to $201$. Like most of ML models, the whole formalism of SVR boils down to a minimization of a loss function that takes on the following form:
\begin{equation}\label{eq:loss}
L = \frac{1}{2}\norm{\hat\Theta}^2 + C\sum^{l} _{i=1}|y_i-f[\rho_i,\hat{\Theta}]|_{\epsilon}.
\end{equation}
$y_i$ is the kinetic energy of the $i^{th}$ sample and $\epsilon$ is the precision parameter of the model, $C$ is a regularization term. The term inside the summation is the Vapnik $\epsilon$ - insensitive error and is given, for one instance, by: 
\begin{eqnarray}
|y_i-f[\rho_i,\hat{\Theta}]|_{\epsilon} = 
\begin{cases}
 0 & \text{if}\ |y_i-f[\rho_i,\hat{\Theta}]|<\epsilon \\ 
 |y_i-f[\rho_i,\hat{\Theta}]|-\epsilon & \text{otherwise}.
\end{cases}
\end{eqnarray}
By introducing the slack variables $\eta_i$ and $\eta^*_i$(shown in Fig. \ref{fig:fig1}) we can write the constrained minimization of the new objective function as the following : 
\begin{eqnarray}\label{eq:optimization}
&&\text{mininize}(\ \frac{1}{2}\norm{\hat{\Theta}}^2 + C\sum^{l}_{i=1}(\xi_i + \xi^*_i))\\
&&\text{subject to}
\begin{cases}
y_i - (\hat{\Theta}^T\rho_i + b) \le\epsilon + \xi_i \\ 
(\hat{\Theta}^T\rho_i + b) - yi \le\epsilon + \xi^*_i\\
\xi_i,\xi^*_i \ge 0,\\
\end{cases}\nonumber
\end{eqnarray}
\noindent where the slack variables $\xi_i$ and $\xi^*_i$ are introduced to account for model errors. By introducing Lagrange multipliers to account for the constraints one ends up with intuitive rather simple expressions 
of the estimator functional and the weights vector, namely: 
\begin{eqnarray}\label{eq:estimator}
&&f[\rho,\hat{\Theta}] = \sum^l_{i=1}(\alpha_i - \alpha^*_i)\langle\ \rho_i,\rho\rangle + b\\ \nonumber
&&\hat\Theta = \sum^m_{i=1}(\alpha_i - \alpha^*_i)\rho_i. 
\end{eqnarray}
\noindent We refer the reader to Ref \cite{Smola} for a detailed passage from Eq. \ref{eq:optimization} to Eq. \ref{eq:estimator}. The dual variables $\alpha_i$ and $\alpha^*_i$ are Lagrange multipliers introduced to account for the first and second constraints in Eq. \ref{eq:optimization}. 
$f[\rho,\hat{\Theta}]$ is the regression function and is given as a linear combination of inner products between the input variable and other training examples. This very simple and sophisticated formula can be extended to non-linear systems (the kind of the current problem) by adopting the kernel trick, whereby the original input density$\rho$ is mapped into a vector $\Psi(\rho)$ from the feature space (a higher dimensional space).
The kernel is given by the inner product between mapped vectors, namely $K(\rho,\rho_i) = \langle\ \Psi(\rho),\Psi(\rho_i)\rangle$. In the new feature space, the dimensionality expansion plays an essential role in rendering the problem linear.
\noindent Throughout this work, we use the Gaussian kernel defined as,
\begin{equation}
K(\rho,\rho_i,\gamma) = \exp(-\frac{\norm{\rho-\rho_i}^2}{2\gamma^2}),
\end{equation}
where $\gamma$ is the variance of the distribution. The feature space estimator can be expressed in terms of $K(\rho,\rho_i,\gamma)$, thus reads: 
\begin{equation}\label{eq:estimatorK}
 \Xi[\rho,\hat\Theta] = \sum^l_{i=1}(\alpha_i-\alpha^*_i)K(\rho,\rho_i,\gamma)+b.
\end{equation}
Moreover, we keep the same expression for the weights vector $\hat\Theta$ whereby we substitute the input vectors by the new feature space mapped vectors, $\Psi(\rho)$:
\begin{equation}\label{eq:weightsK}
\hat\Theta = \sum^l_{i=1}(\alpha_i - \alpha^*_i)\Psi(\rho_i).
\end{equation}

\section{Kinetic Energy Calculation}\label{sec:sec3}

\subsection{Non linearity of the problem} 

For linear problems, one can use Eq. \ref{eq:estimator} to build the estimator, however, nonlinear systems (like the one studied here) demand passing through a high dimensional space where the new estimator and the weights can be expressed in terms of kernels and mapping functions as given in Eq. \ref{eq:estimatorK} and \ref{eq:weightsK}, respectively. In Fig. \ref{fig:fig1}, we used SVR in order to calculate the kinetic energy as function of the densities principal components (PCs)(Fig. \ref{fig:fig1}-c) and kernel-principal components (KPCs)(Fig. \ref{fig:fig1}-d). A two fold cross validation with linear regression has been used for training purpose. As expected, the increase of the number of PCs (or KPCs) reduces the mean absolute error (MAE) in the KE calculation. While the error calculation in (c) saturated to the value $0.27$ HA, the error in (d) headed to a very small value ($0.008$ HA), an evidence of a readily non-linear problem. 

\begin{figure}[h!]
  \centering
  \includegraphics[width=9.cm]{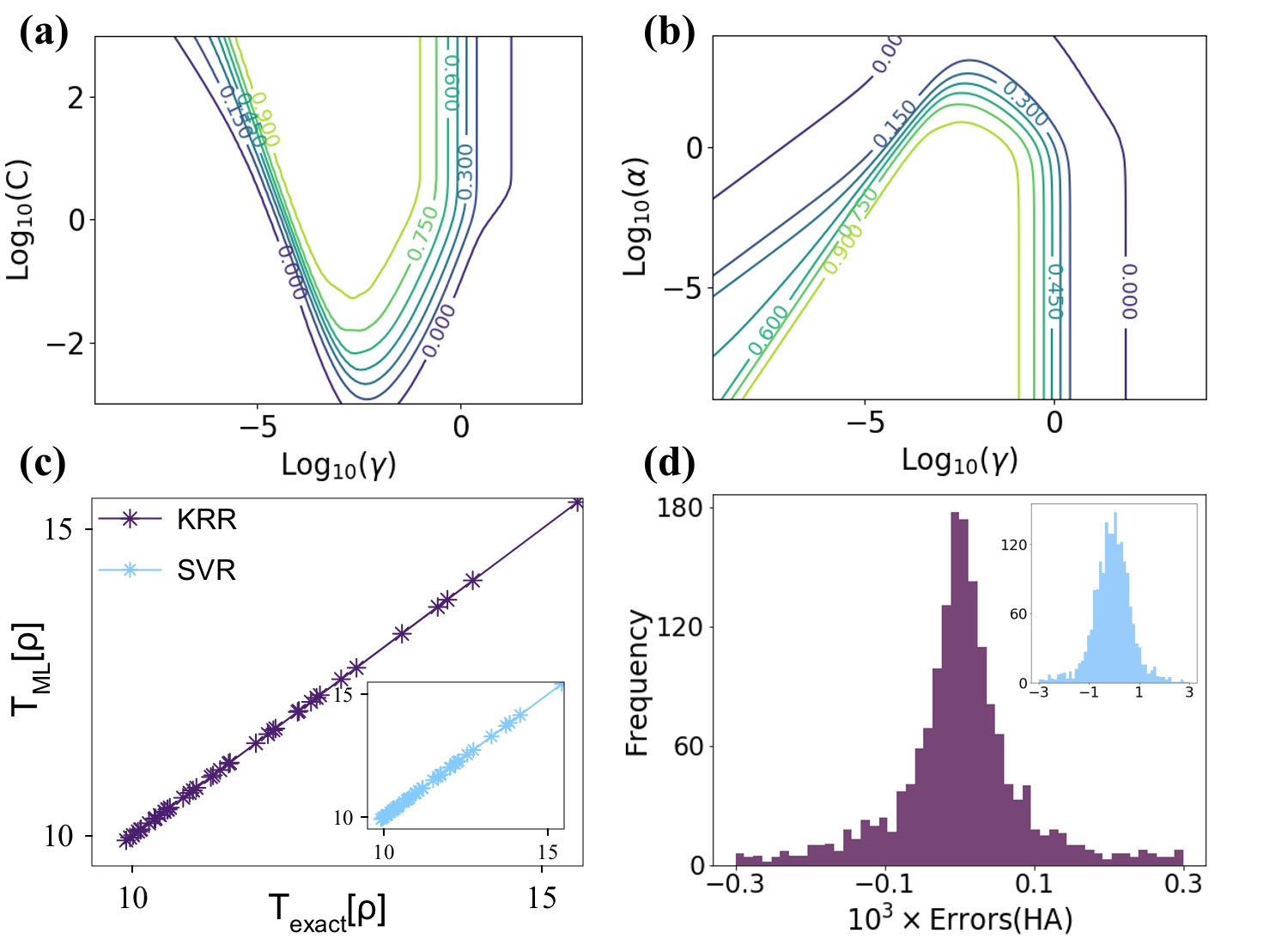}
  \caption{(Color online) (a) and (b) are contour plots showcasing the hyperparameter optimization for SVR and KRR, respectively. (c) (inset of (c)) is a  		parity plot comparing exact values of the KE functional computed using Galerkin method against predictions based on KRR (SVR). (d) (inset 		of (d)) is a histogram showing the quasi-normal distribution of the KE estimation errors using KRR (SVR). }\label{fig:fig2}
\end{figure}

As a matter of fact, the established expressions of the KE for $N=1$ (Eq. \ref{eq:Weizsacker}) and $N \rightarrow \infty$ (Eq. \ref{eq:Thomas}) pour into a non-linear character of the functional. Insets of (c) and (d) are 3 dimensional visualizations of the MAE as function of the first two principal components (containing most of the variance) without and with applying the kernel, respectively. Inset of (c) shows a compelling 3 dimensional behavior while a tendency of a dimensionality reduction can be grasped in the inset of (d). In (c) MAE reached a plateau of constant error after including 7 PCs. This number increased after applying the kernel to 20 K-PCs. Indeed, the reduction of the problem complexity comes with the cost of dealing with high dimensional space, hence, a trade off between these two needs to be checked carefully. This trade-off can be controlled by means of the precision parameter $\epsilon$ which is tightly related to the number of support vectors involved in the optimization process of Eq. \ref{eq:optimization}. Moreover, it is clear that the cross-validation error measured after the application of the kernel is $1$ to $2$ orders of magnitude smaller than the error with no kernelization. That is due to the non-linear character of the KE functional which cannot be modeled accurately using simple linear regression. Subsequently, we used the (kernel - based) SVR to train our non-linear model. During the training, the minimization of the loss function (see Eq. \ref{eq:loss})  results in the obtention of the Lagrange multipliers $\alpha_i$ and $\alpha^*_i$ which, in their turn, are used to estimate the kinetic energy according to Eq. \ref{eq:estimator}. Saying that, one needs to choose carefully a different category of parameters which are not included in the optimization of the loss function, are instead user defined. These hyperparameters can be chosen by cross-validating the training data and sorting the corresponding MAE. 

\begin{figure*}
  \centering
  \includegraphics[width=13cm]{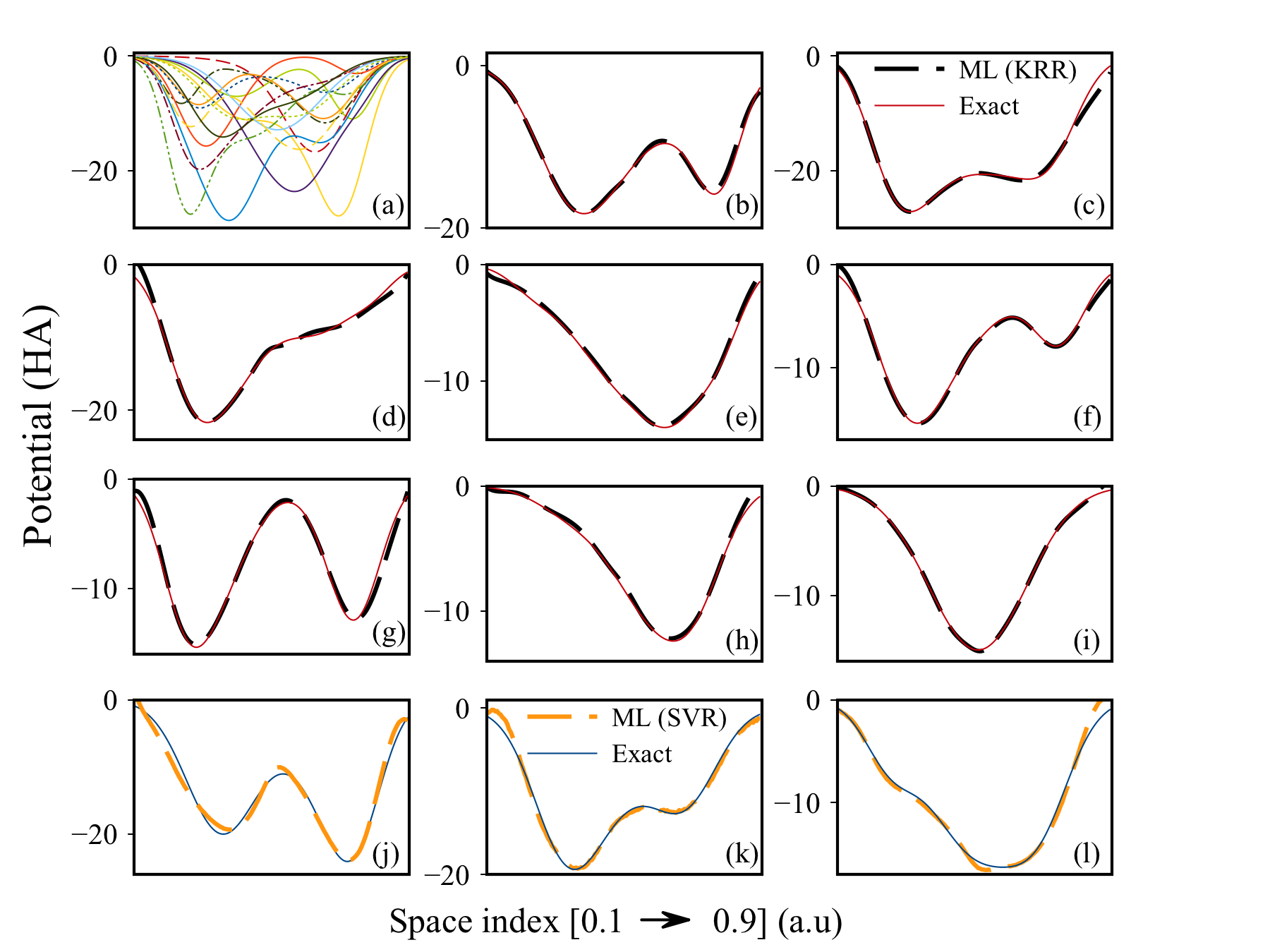}
  \caption{(Color online) Exact (solid lines) and machine learning calculated potentials (dashed lines). (a) presents a random set of potentials used to generate electron densities for the training purpose. The first three rows are potentials calculated using KRR for which different plots correspond to potentials with different number of dips. The last row corresponds to results obtained using the SVR model.}  \label{fig:fig3}
\end{figure*}

\subsection{Hyperparameters optimization and Kinetic Energy calculation}

Figure \ref{fig:fig2}(a-b), depicts contour plots of the KE estimation scores as function of $\gamma$ and $C$ for the SVR model, and of $\gamma$ and $\alpha$ for the KRR model. In both plots $\gamma$ is a measure of the kernel distribution variance. $C$ is a regularization term that penalizes the model error in SVR and $\alpha$ is a regularization term that weights the $L2$ norm term in KRR. A decimal logarithmic scale is applied in both plots. 
In Fig. \ref{fig:fig2}(a), an increase of $C$ favors higher scores (less estimation errors on the cross-validation set) which overpassed $0.9$ for $C>10^{-1}$. The optimal values of $\gamma$ spans the range between $10^{-5}$ and $10^{-1}$.
In a similar fashion, the optimal range of the KRR hyperparameters coincides with $\gamma$ spanning the interval of values less than $1$. A small value of $\alpha$ leads to a better estimation score as $\alpha$ plays the role of the inverse of the regularizer term in other regression based ML algorithms (like $1/2C$ in the case of logistic regression). Supported by the data of Fig. \ref{fig:fig2}, we choose $\epsilon$ to be equal to $5\times10^{-4}$,  $C=100$ and $\gamma=5\times10^{-3}$ for SVR, $\alpha$ and $\gamma$ take on the values $10^{-5}$ and $0.07$, respectively for the KRR. Fig. \ref{fig:fig2}((c) and (d)) are necessary conditions for an accurate (and generalizable) ML-based estimator of the KE functional. In the parity plot (c), kinetic energies of $40$ instances of the test set (for potentials with $5$ dips) have been reported. The curve points (red spheres) have the ML(KRR) - calculated KE as ordinate and the real (calculated using Galerkin method) KE as abscissa.
From first glance, a good match between the KE calculated using both methods appear to take place. This plot is associated with a mean absolute error, MAE $=1.6\times10^{-4}$ HA, and a relative mean absolute error, RMAE $=1.4\times10^{-5}$ HA. Similar results are obtained using SVR (inset of (c)) with MAE $=5\times10^{-4}$ HA and RMAE $=4.4\times10^{-5}$ HA. Fig. \ref{fig:fig2}(d)(inset of (d)) represents the distribution of errors, calculated using KRR (SVR), for 1900 samples. Both distributions are nearly normal and centered around the value $0$.
Note that normal distribution of errors reflects a distribution with no pattern, a necessary condition which insures the generalizability of the model. Overall, the reached accuracy, the non skewed distribution of errors and the perfect parity plot constitute the cornerstone of approaching OFDFT with ML. Nevertheless, calculating the KE itself is not the main challenge addressed in the current work (as it has been addressed in previous work\cite{SnyderPRL}), but being able to calculate its functional derivative in a direct way and avoiding complexity resulting from using indirect calculation schemes. Saying that, a careful treatment of these derivatives needs to be applied in order to reduce errors in calculating the total energy. 

\section{Direct Scheme Derivative Calculation}\label{sec:sec4}

In order to calculate the derivative of the KE functional $\Xi[\rho]$, we express $\Xi[\rho]$ as a Taylor series expansion\cite{Dreizler}:
\begin{equation}\label{eq:func_taylor}
\Xi[\rho+\eta \phi]=\sum^{N}_{n=0}\frac{d^n\Xi[\rho+\eta \phi]}{d\eta^n}\bigg\rvert_{\eta=0} \eta^n + O(\eta^{N+1}).
\end{equation}

\begin{figure}[h!]
  \centering
  \includegraphics[width=9.cm]{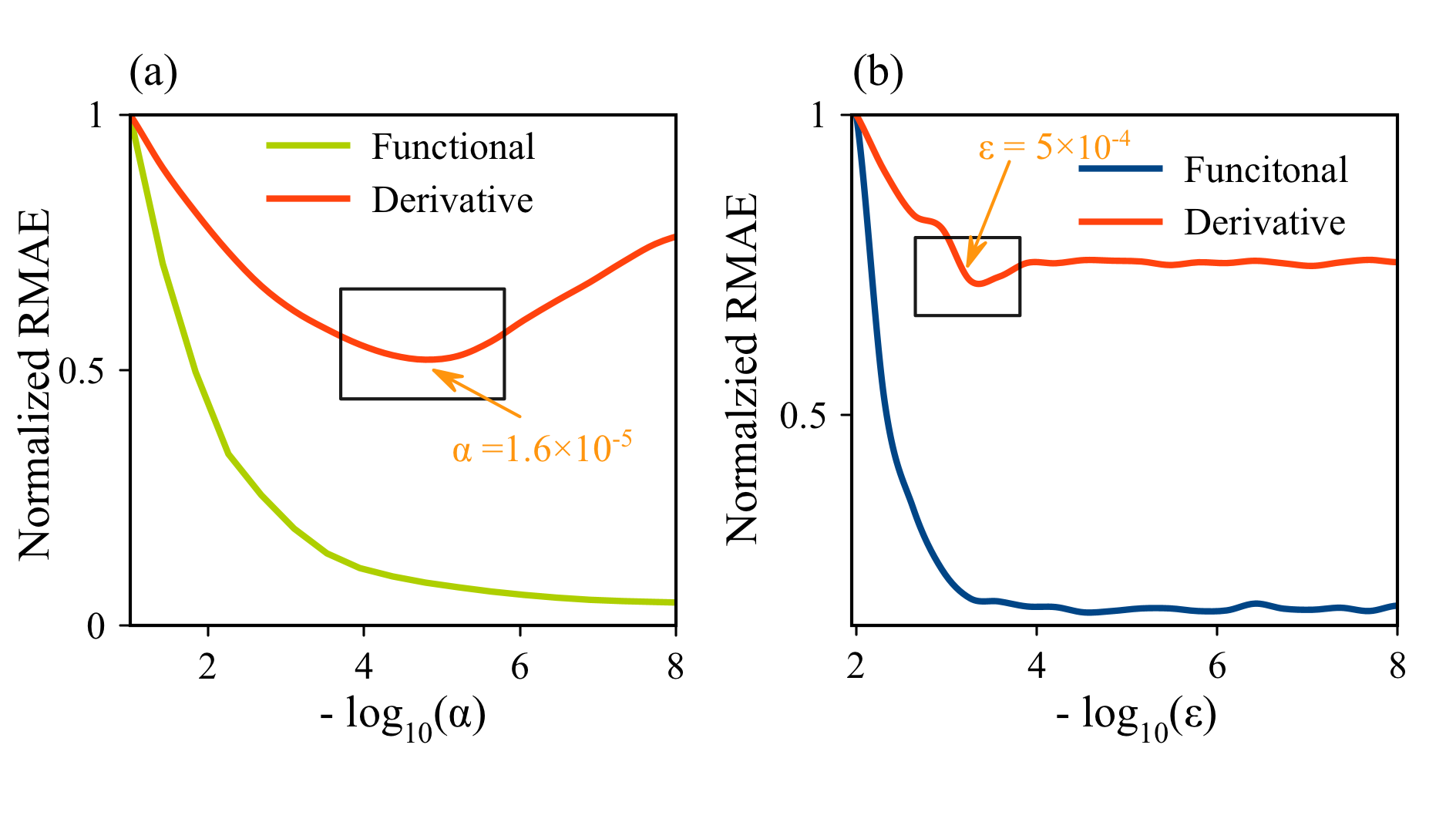}
  \caption{Variation of the normalized mean absolute error obtained while calculating the KE functional (green and blue) and its derivative (red) as function of accuracy parameters of the ML tool. (a) is the result of the KE estimation of $100$ samples of unseen data using KRR. (b) is same as (a) but using SVR.} \label{fig:fig4}
\end{figure}

Then, the functional derivative can be expressed in the following way\cite{Parr}:
\begin{equation}\label{eq:func_derv}
\lim_{\eta \rightarrow 0}{\frac{\Xi[\rho+\eta \phi]-\Xi[\rho]}{\eta}}=\frac{d}{d\eta}(\Xi[\rho+\eta \phi])\bigg\rvert_{\eta=0},
\end{equation}
where $\Xi$ is the KE functional to be derived, $\eta$ is a an infinitesimal coefficient and $\phi$ is a random function. We use the left hand side of Eq. \ref{eq:func_derv} as a numerical implementation of the mathematical formula.
A delta distribution $\delta$ is used instead of $\phi$ to account for the variation of the kinetic energy functional, such a choice is seen to be mathematically correct but also handy for our numerical implementation.\\
Fig. \ref{fig:fig3}  showcases the external potential calculated from machine learning by differentiating the kinetic energy estimator the way dictated by Eq. \ref{eq:func_derv}. The real potentials are present for reference. 
\noindent Each three consecutive plots (from left to right) designate same number of dips of the original potential with random depth, center and spread ($a$, $b$ and $c$ in Eq. \ref{eq:potential}). As has been mentioned in the preceding text, the functional differentiation is done without any additional training. 
Although the accuracy of the original KE estimator is not enough to account for the infinitesimal parameter $\eta$, the use of a proper standardization of the input data reduces the discrepancy between the density $\rho$ and the infinitesimal variation $\eta \phi$ (in Eq. \ref{eq:func_derv}), making the evaluation of $\rho + \eta\phi$ falling within the range of accuracies tolerated by the estimator. Here we choose $\eta$ to be equal to $5\times10^{-8}$.
It is worth noting that in some cases, larger discrepancies might be detected between the ML calculated potentials and the real ones. These mostly happen at flat surfaces of the potential and might be caused by inevitable numerical errors while evaluating the functionals derivatives. The first three rows of Fig. \ref{fig:fig3} are results obtained using the KRR algorithm while the remaining row showcases the derivatives obtained using SVR for training the kinetic energy functional. The training data set used to train the KE estimator was a mixture of potentials with random parameters and with different number of dips. We believe that this choice was an enhancing factor which led to more generalizable KE estimator.
Most of the plots in Fig. \ref{fig:fig3} reveal a good match between real and ML calculated potentials. Arguably, the main reasons behind the direct obtention of the derivatives are twofold. In one hand, we have enriched the distribution of the training examples by introducing more general potentials (compared to potentials used in similar works\cite{Fahhad2017,SnyderPRL}) with dips randomly positioned (see Fig. \ref{fig:fig3}(a)). At times, these potentials might not be physically realistic, however, they surely favor the generalizability character of our method.
From ML point of view, the regularization was essential to prevent the overfitting of the model. Together with the regularization, the standardization plays a decisive role in bringing the infinitesimal change of the density falling in a range tolerated by the KE estimator accuracy.   

\subsection*{Impact of  the Regularization}

We define a mean absolute error metric for the KE derivative denoted RMAE (for relative mean absolute error).  RMAE is defined as an error averaged over the space grid values of the derivative for each example and over the whole test set examples.
 \begin{equation}
 RMAE=\frac{1}{l_t n_x}\sum^{l_t}_{j=1}\sum^{n_x}_{i=1}{\abs{\frac{\delta \Xi[\rho,\hat\Theta]}{\delta \rho_j(x)}-\frac{\delta T[\rho]}{\delta \rho_j(x)}}_{x_i}}
 \end{equation}

where $l_t$ and $n_x$ denote the size of the test set and the grid space dimension, respectively. 

Figure. \ref{fig:fig4} displays the variation of the relative mean squared errors of both the kinetic energy functional and its derivative - estimators. In Fig. \ref{fig:fig4}(a) RMAE calculations are performed using KRR method, while the results of applying the SVR method are illustrated in (b). The errors variations are plotted against the main precision enhancing parameters of both tools, $\alpha$ for KRR and $\epsilon$ for SVR. The logarithmic scale is used for the abscissa of both plots for the sake of readability. Ordinates are normalized to the value with smallest precision parameters bringing all errors plots to comparable scales. Using both ML tools, the variation of the KE functional RMAE showcases a strictly monotonic behavior in \ref{fig:fig4}(a), where RMAE decreases with increasing $\alpha$. \\
\noindent In spite of the slightly staggered behavior in \ref{fig:fig4}(b) for larger $\epsilon$, these small variations does not prohibit identifying the regime as strictly monotonic. In contrary, the derivative RMAE profiles show, in overall, two identifiable regimes. A first descending curve with increasing $\alpha$($\epsilon$) up to a minimal value $\alpha=1.6\times10^{-6}$($\epsilon=5\times10^{-4}$) is seen in \ref{fig:fig4}(a) (\ref{fig:fig4}(b)), respectively. The second regime corresponds to values of $\alpha>1.6\times10^{-4}$ and $\epsilon>5\times10^{-4}$, where the curves witness a monotonic increase. As can be denoted from both figures, these intervals of increase of the functional derivative errors are associated with a decrease of the KE functional errors. As a matter of fact, This dissimilarity between both errors behaviors is caused by what is coined as overfitting. The increase of the KE functional accuracy comes with the cost of overfitting the data, i.e, being capable of predicting kinetic energies with high accuracy but not being generalizable to the derivative case. Moreover, RMAE increased drastically for very small values of $\alpha$ (ascending direction for KE estimator accuracy), making the prediction of the functional derivative prone to errors. Here, the similar behavior of both estimators proves the universality of this insight despite the use of different tools with different parameters thresholds.

\section{Conclusion}\label{sec:sec5}

In conclusion, we have shown that the derivative of the KE estimator can be calculated, in a direct way, without any further training (except for obtaining the KE estimator). Throughout this work, we have emphasized the importance of a careful choice of the ML regularized parameters. In spite of its role in choosing the right predictor, the accuracy must be treated as a double-edged criteria (see Fig. \ref{fig:fig4}). 
The model built in the current work enhances the applicability of machine learning techniques applied to density functional theory and gives a solution to a direct calculation of the functional derivatives. It can be further ameliorated by considering datasets of more general physical systems or/and by applying numerical techniques that can account for  physical constraints.

\end{document}